\documentclass[a4paper,11pt]{article}
\usepackage{pos}
\usepackage[export]{adjustbox} 
\title{Hyperon Pair Production at BESIII}

\author*[a,b]{Zijie Shang}
\author[a,b]{Ruoyu Zhang}
\author[a,b]{Xiongfei Wang}
\onbehalf{on behalf of BESIII Collaboration}

\affiliation[a]{School of Physical Science and Technology, Lanzhou University, Lanzhou 730000, People’s Republic of China}

\affiliation[b]{Lanzhou Center for Theoretical Physics,
Key Laboratory of Theoretical Physics of Gansu Province,
Key Laboratory of Quantum Theory and Applications of MoE,
Gansu Provincial Research Center for Basic Disciplines of Quantum Physics, Lanzhou University, Lanzhou 730000, People’s Republic of China}

\emailAdd{shangzj18@lzu.edu.cn}

\abstract{
This poster presents recent experimental findings from the BESIII experiment, including two categories of studies. Firstly, it reports the studies on the hyperon pair production in the processes $e^+e^- \to \Lambda\bar\Lambda$, $\Sigma^+\bar\Sigma^-$, $\Sigma^0\bar\Sigma^0$, $\Xi^0\bar\Xi^0$ and $\Xi^-\bar\Xi^+$. The Born cross sections for these processes are measured at center-of-mass energies ranging from $3.5$ to $4.9$ GeV. For the first time, evidence for the decays $\psi(3770)\to\Lambda\bar\Lambda$ and $\Xi^-\bar\Xi^+$ is found. Secondly, it reports the measurements of the Born cross sections of multi-body decays involving hyperon final states. The evidence of $\psi(4160)\to K^-\bar\Xi^+\Lambda+c.c.$ is found. These results offer new perspectives on hyperon production in $e^+e^-$ annihilation and the decay mechanism of vector charmonium(-like) states, contributing to our understanding of hadron dynamics. 
}

\FullConference{
The 21st International Conference on Hadron Spectroscopy and Structure (HADRON2025)
}

\begin{document}
\maketitle

\section{Introduction}
Study of hadron production in \(e^+e^-\) annihilation above the open charm threshold is crucial for understanding the nature of charmonium(-like) states and testing the non-perturbation theory of QCD. The overabundance of the vector-charmonium(-like) states and the discrepancies between the potential model predictions and experimental measurements provide a significant opportunity to probe exotic configurations of quarks and gluons. Many theoretical models, such as hybrid, multiple-quark state, and molecule, etc., have been proposed to interpret the charmonium(-like) states~\cite{Chen:2016qju}. However, no solid conclusion has yet emerged and the true nature of these states remains a puzzle. This status reflects our poor understanding of the behavior of the strong interaction in the non-perturbative regime. To make progress, more high-precision measurements are required. Among these measurements, studies of the hyperon decays of charmonium(-like) states hold particular promise due to the simple topologies of the final states and relatively well-understood mechanisms. Additionally, the discovery of charmonium(-like) states in \(e^+e^-\) annihilation into charmonium and light hadrons highlights the importance of studying hyperon final states where information is still limited~\cite{Wang:2024gtk}. 

\section{The BEPCII/BESIII experiment}
The BESIII detector~\cite{Ablikim:2009aa} records symmetric $e^+e^-$ collisions provided by the BEPCII storage ring~\cite{Yu:IPAC2016-TUYA01} in the center-of-mass energy range from 1.84 to 4.95 GeV, with a peak luminosity of $1.1 \times 10^{33}\;\text{cm}^{-2}\text{s}^{-1}$ achieved at $\sqrt{s} = 3.773\;\text{GeV}$. BESIII has collected large data samples in this energy region~\cite{Ablikim:2019hff,EcmsMea,EventFilter}. The cylindrical core of the BESIII detector covers 93\% of the full solid angle and consists of a helium-based multilayer drift chamber~(MDC), a time-of-flight system~(TOF), and a CsI(Tl) electromagnetic calorimeter~(EMC), which are all enclosed in a superconducting solenoidal magnet providing a 1.0~T magnetic field (0.9~T in 2012). The solenoid is supported by an octagonal flux-return yoke with resistive plate counter muon identification modules interleaved with steel. The acceptance of charged particles and photons is 93\% over the $4\pi$ solid angle. The charged-particle momentum resolution at $1~{\rm GeV}/c$ is $0.5\%$, and the ${\rm d}E/{\rm d}x$ resolution is $6\%$ for electrons from Bhabha scattering. The EMC measures photon energies with a resolution of $2.5\%$ ($5\%$) at $1$~GeV in the barrel (end cap) region. The time resolution in the plastic scintillator TOF barrel region is 68~ps, while that in the end cap region was 110~ps.  The TOF end cap system was upgraded in 2015 using multi-gap resistive plate chamber technology, providing a time resolution of 60~ps, which benefits the data quality~\cite{etof1,etof2,etof3}.

\section{Recent advances}
\subsection{Two body decay}
\begin{figure}[!htbp]
\begin{center}
\centering
\includegraphics[width=0.98\textwidth, valign=c]{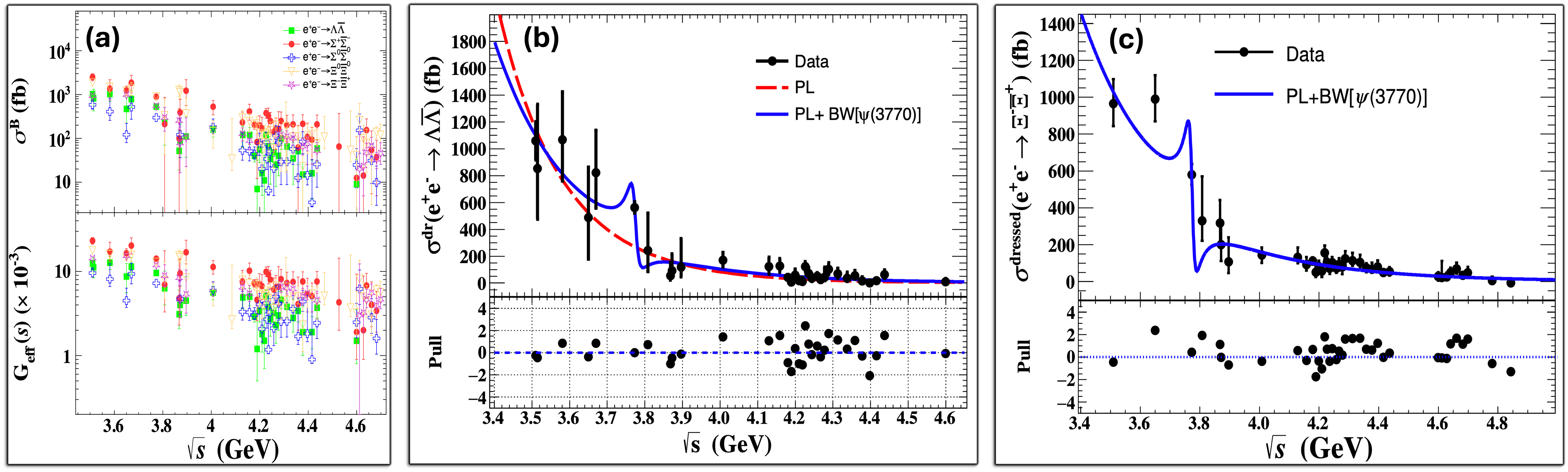}

\end{center}
\caption{\label{twobody}Comparison of Born cross sections and effective form factors as a function of CM energy for different decay channels (Left). Fits to the dressed cross sections for $e^+e^- \to \Lambda\bar{\Lambda}$ (Middle) and $e^+e^-\to\Xi^-\bar\Xi^+$ (Right). }
\end{figure}

The BESIII experiment performed its first search for the charmonium(-like) states $Y(4230/4260)$ in the process $e^+e^- \to \Xi^- \bar{\Xi}^+$ at center-of-mass energies ranging from 3.5 to 4.6 GeV~\cite{XiXiprl}. No significant evidence of $Y(4230/4260)$ decaying into the $\Xi^-\bar{\Xi}^+$ final states was observed. Subsequently, BESIII has conducted a series of studies on hyperon pair production in the processes $e^+e^- \to \Lambda\bar{\Lambda}$, $\Sigma^+\bar{\Sigma}^-$, $\Sigma^0\bar{\Sigma}^0$, $\Xi^0\bar{\Xi}^0$, and $\Xi^-\bar{\Xi}^+$~\cite{LamLam, SigSigc, Sig0Sig0, Xixi0, XiXiimp}. The Born cross sections and effective form factors for these processes were measured, as shown in Fig.~\ref{twobody}(a). To further investigate the charmonium(-like) states, fits to the dressed cross sections were carried out. Two pieces of evidence for the decays $\psi(3770)\to \Lambda\bar\Lambda$ and $\Xi^-\bar\Xi^+$ were found, as shown in Fig.~\ref{twobody}(b) and (c). For other processes, no significant signal of charmonium(-like) states was found. Only the upper limits for the product of the branching fraction and the electronic partial width $\Gamma_{ee}{\cal{B}}$ at the 90\% confidence level (C.L.) were evaluated. These measurements provide evidence for charmless decays of $\psi(3770)$ and can be useful for understanding the charmonium(-like) states coupling to the hyperon pair final states.

\subsection{Multi-body Decay}
\begin{figure}[!htbp]
\begin{center}
\centering
\includegraphics[width=0.75\textwidth, valign=c]{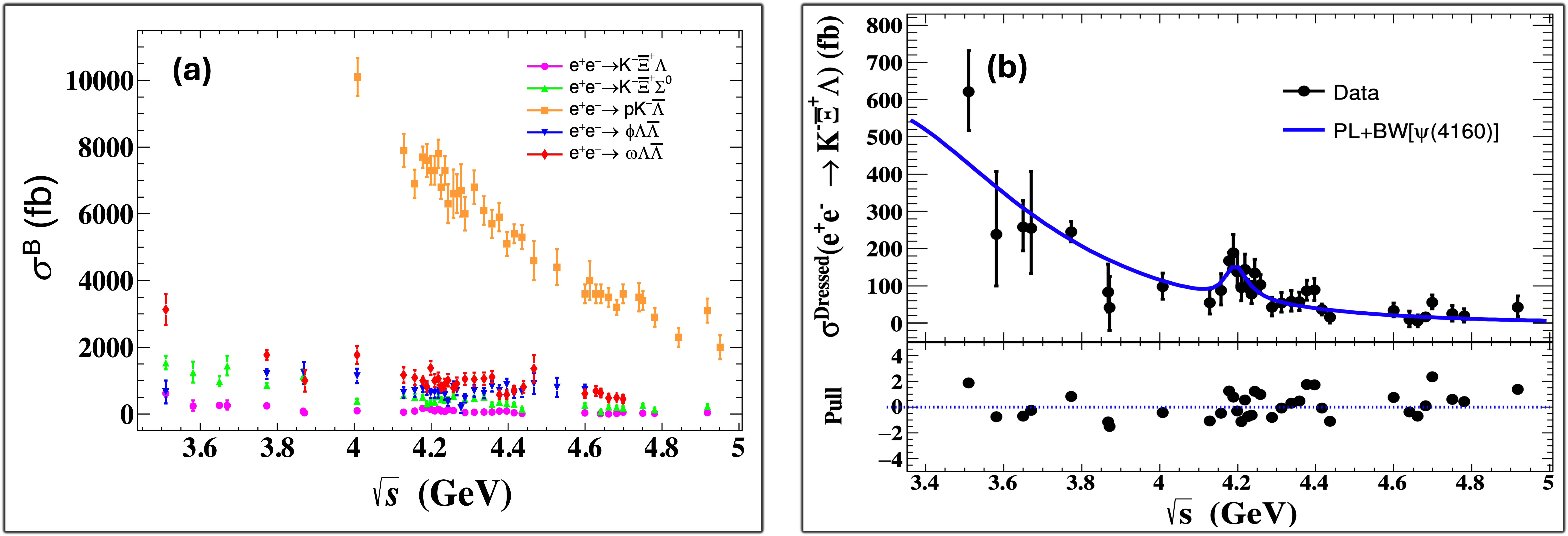}
\end{center}
\caption{\label{multibody}Comparison of Born cross sections as a function of CM energy for different decay channels (Left). Fit to the dressed cross section with the assumption of $\psi(4160)$ resonance plus a continuum contribution for the decay of $e^+e^- \to K^-\bar\Xi^+\Lambda$ (Right).}
\end{figure}

The BESIII experiment has also reported measurements of the Born cross sections for multi-body decays involving hyperon final states~\cite{KXiLS, LLeta, LLphi, pkl}, as shown in Fig.~\ref{multibody}(a). For these processes, no significant charmonium(-like) state was found, except for an evidence for the decay of $\psi(4160)\to K^-\bar\Xi\Lambda+c.c.$ with a significance of 4.4 $\sigma$ including systematic uncertainty, as shown in Fig.~\ref{multibody}(b). The upper limits for $\Gamma_{ee}{\cal{B}}$ at the 90\% C.L. for assumed resonances decaying into the $K^-\bar{\Xi}^+\Lambda/\Sigma^0$ final state were determined. These results are valuable as they contribute to the experimental information regarding the three-body baryonic decay of charmonium(-like) states. 

\begin{figure}[!htbp]
\begin{center}
\centering
\includegraphics[width=0.75\textwidth, valign=c]{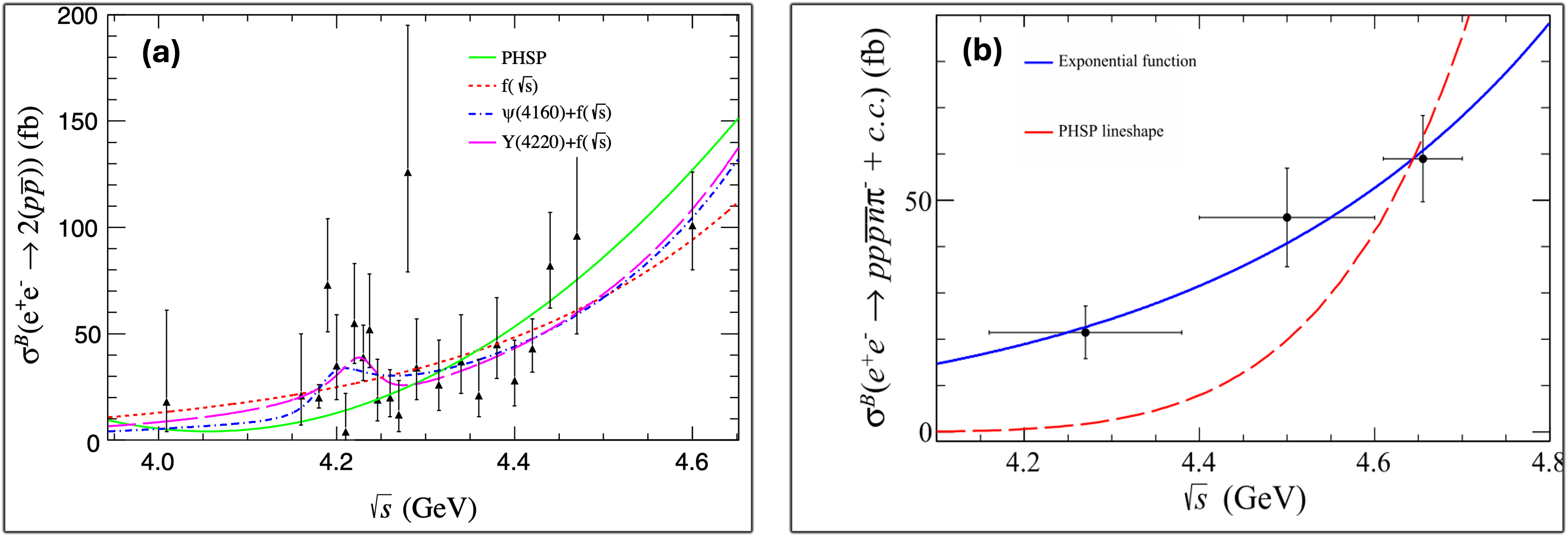}
\end{center}
\caption{\label{45body}Born cross sections of the process $e^+e^-\to 2(p\bar p) $ as a function of CM energy (Left) and the average Born cross sections at three sub-samples for the process $e^+e^-\to pp\bar p\bar n\pi^-+c.c.$}
\end{figure}

In addition, the BESIII experiment has searched for charmonium(-like) states through $e^+e^-\to 2(p\bar p)$~\cite{pppp} and $e^+e^-\to pp\bar p\bar n\pi^- + \text{c.c.}$~\cite{pppnpi}. Fig.~\ref{45body}(a) illustrates the measured Born cross sections, and the lineshape can generally be described by an empirical exponential function. The significance of possible contributions from a $\psi(4160)$ or $Y(4220)$ resonance is small, namely 0.83 $\sigma$ and 1.69 $\sigma$, respectively. With the present statistics, it is impossible to draw any conclusion regarding whether there are actual resonances or structures in this lineshape. The average Born cross sections for the decay of $e^+e^-\to pp\bar p\bar n\pi^- + \text{c.c.}$ in three energy ranges were measured as shown in Fig.~\ref{45body}(b), revealing discrepancies with the expectation from five-body phase space (PHSP). However, due to the limited statistics, no significant charmonium(-like) state was found. 

\section{Summary}
The BESIII experiment has been operating successfully since 2008 and has collected the largest data samples in the $\tau$-charm physics region. Many advances in hadron production in $e^+e^-$ annihilation at BESIII have been achieved. These include the evidences of $\psi(3770)\to\Lambda\bar\Lambda/\Xi^-\bar\Xi^+$ and $\psi(4160)\to K^-\bar\Xi^+\Sigma^0 +c.c.$, among others. Additionally, searches in $e^+e^-\to 2(p\bar p) $ and $e^+e^-\to pp\bar p\bar n\pi^-+c.c.$ have been performed. However, no significant signal was observed due to limited statistics. These findings offer valuable insights into the production of hyperon final states, contributing to our understanding of hadron dynamics within this energy regime and shedding light on the mechanisms and properties of charmonium(-like) states. Thus, they enhance our comprehension of hadronization dynamics and quark-gluon interactions. 

\section{Acknowledgment}
This work is supported in part by
the Fundamental Research Funds for the Central Universities under Contracts Nos. lzujbky-2025-ytA05, lzujbky-2025-it06, lzujbky-2024-jdzx06;
National Natural Science Foundation of China under Contracts No. 12247101; 
the Natural Science Foundation of Gansu Province under Contracts Nos. 22JR5RA389, 25JRRA799; 
the 111 Project under Grant No. B20063.

\end{document}